\documentclass[a4paper]{article}
\usepackage{epsfig,slashed}
\begin{document}
\textwidth=135mm
 \textheight=200mm
\begin{center}
{\bfseries 
Bound states and superconductivity in dense Fermi systems}
\vskip 5mm
D. Blaschke$~^{1,2}$ and D. Zablocki $~^{1,3}$
\vskip 5mm
{\small 
{\it $~^1$ Instytut Fizyki Teoretycznej, 
Uniwersytet Wroc{\l}awski, 50-204 Wroc{\l}aw, Poland}} 
\\
{\small 
{\it $~^2$ Bogoliubov Laboratory for Theoretical Physics,
JINR, 141980 Dubna, Russia}}
\\
{\small 
{\it $~^3$ Institut f\"ur Physik der Universit\"at, 
D-18051 Rostock, Germany}} 
\end{center}
\vskip 5mm
\centerline{\bf Abstract}{ 
A quantum field theoretical approach to the thermodynamics of dense Fermi 
systems is developed for the description of the formation and dissolution of 
quantum condensates and bound states in dependence of temperature and density.
As a model system we study the chiral and superconducting phase transitions 
in two-flavor quark matter within the NJL model and their interrelation with 
the formation of quark-antiquark and diquark bound states.
The phase diagram of quark matter is evaluated as a function of the diquark
coupling strength and a coexistence region of chiral symmetry breaking and 
color superconductivity is obtained at very strong coupling.
The crossover between Bose-Einstein condensation (BEC) of diquark bound states 
and condensation of diquark resonances (Cooper pairs) in the continuum (BCS) 
is discussed as a Mott effect. 
This effect consists in the transition of bound states into the continuum of 
scattering states under the influence of compression and heating. 
We explain the physics of the Mott transition with special emphasis on role 
of the Pauli principle for the case of the pion in quark matter. 
\vskip 5mm
}
\section{\label{sec:intro}Introduction}{
Key issues of modern physics of dense matter are concepts explaining 
the phenomena related to the appearance of 
quantum condensates in dense Fermi systems.
Two regimes are well-known: the Bose-Einstein condensation (BEC) of 
bound states with an even number of Fermions and the condensation of 
bosonic correlations (e.g., Cooper pairs) in the continuum of unbound states 
according to the Bardeen-Cooper-Schrieffer (BCS) theory.
While the former mechanism concerns states which are well-localized in 
coordinate space as they occur for strong enough attractive coupling, the 
latter mechanism applies to states which are correlated 
within a shell of the order of the energy gap $\Delta$ around the    
Fermi sphere in momentum space but delocalized in coordinate space. 
The transition between both regimes is called BEC-BCS crossover.
Recently, this transition regime became accessible to laboratory experiments
with ultracold gases of fermionic atoms coupled via Feshbach resonances
with a strength tunable by applying external magnetic fields, see Fig. 1.
\begin{figure}[htb]
\epsfig{figure=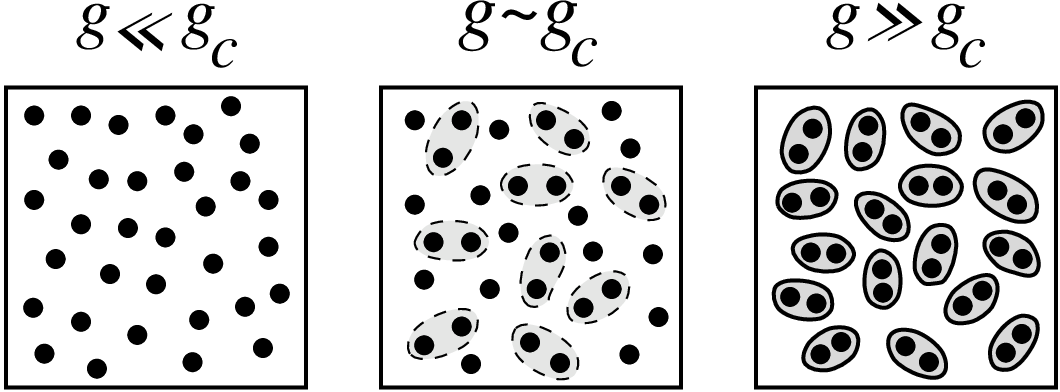,width=0.9\textwidth,angle=0}
\caption{Illustration of the transition from BCS pairing with delocalized 
wave functions to BEC of bound states, well-localized in coordinate space,
from Ref. \cite{chenfig}}
\end{figure} 
After the preparation of fermionic dimers in 2003, 
now also the BEC \cite{greiner:2003,Zwierlein:2003} and 
superfluidity of these dimers has been observed 
\cite{Zwierlein:2005,Greiner:2003a}.
The BEC-BCS crossover is physically related \cite{Calzetta:2006}
to the bound state dissociation or Mott-Anderson delocalization transition 
\cite{Mott} where the modification of the effective coupling strength is 
caused by electronic screening and/or Pauli blocking effects \cite{Ebeling:2008}.
It is thus a very general effect expected to occur in a wide variety of dense 
Fermi systems with attractive interactions \cite{Armen:2006} such as 
electron-hole systems in 
solid state physics \cite{Bronold:2006}, electron-proton systems in the 
interior of giant planets \cite{Redmer:2006}, deuterons in nuclear matter 
\cite{Schmidt:1990
} or diquarks in quark matter 
\cite{Kitazawa:2001ft
}.
The BEC-BCS crossover transition in quark matter is of particular theoretical 
interest due to the additional relativistic regime 
\cite{Abuki:2006dv
}.

A systematic treatment of these effects is possible within the path integral
formulation for finite-temperature quantum field theories.
This approach is rather general as it is relativistic and is especially suited 
to take into account the effects of spontaneous symmetry breaking.
Within these lectures we will present the basics of this approach on the 
example of a model field theory of the Nambu--Jona-Lasinio type for a 
relativistic strongly interacting Fermi system. 
These investigations are also motivated by the analogies of the strongly 
coupled quark-gluon plasma (sQGP) at Relativistic Heavy Ion Collider (RHIC) in  
Brookhaven \cite{Shuryak:2006ap} with the experiments on BEC of atoms in traps.
Furthermore, qualiative insights into possible effects observable in the 
upcoming CBM experiment at FAIR Darmstadt as well as from neutron stars with
quark matter interiors could be derived along the lines of this approach.
}
\section{Path integral formalism}{
}
\subsection{Partition function and model Lagrangian}{
As a generic model system for the description of hot, dense Fermi-systems with 
strong, short-range interactions we consider quark matter described by a 
model Lagrangian with four-fermion coupling.
The key quantity for the derivation of thermodynamic properties is the 
partition function $\mathcal{Z}$ from which all 
thermodynamic quantities can be derived.
It is given as a path integral which in the imaginary time formalism  
(${t=-i\tau}$) can be expressed as \cite{Kapusta:1989}
\begin{equation}
\mathcal{Z} =
\int \mathcal{D}(iq^\dagger)\mathcal{D}(q)\;
e^{\int^\beta \,d^4x\,(\mathcal{L}-\mu q^{\dagger}q)}~,
\label{LL}
\end{equation}
where the chemical potential $\mu$ is introduced as a Lagrange multiplier for
assuring conservation of baryon number as a conserved charge carried by the
quarks. 
The notation $\int^\beta\, d^4x$ is shorthand for 
$\int_0^\beta\,d\tau \int\,d^3x$ where $\beta = 1/T$ is the inverse 
temperature.
The quark matter is described by a Dirac Lagrangian with internal degrees of 
freedom ($N_f = 2$ flavors , $N_c = 3$ colors), with a current-current-type 
four-fermion interaction inspired by one-gluon exchange
\begin{eqnarray}
\label{eq:lag}
{\mathcal L}=\bar{q}(i\slashed\partial - m_0)q
                     -\frac{g^2}{2}\sum_{a=1}^8
                     \bar{q}\frac{\lambda^a}{2}\gamma_\mu q
                     \,\bar{q}\frac{\lambda^a}{2}\gamma^\mu q~,                
\end{eqnarray}
where $\lambda^a$ are the Gell-Mann matrices for color $SU(3)$. After
Fierz transformation of the interaction, we select scalar diquark channel 
and the scalar and pseudoscalar channel so that our model 
Lagrangian assumes the form
\begin{eqnarray}
\mathcal{L} = \mathcal{L}_0 + \mathcal{L}_{qq} + \mathcal{L}_{q\bar{q}} 
\end{eqnarray}
where the different terms are given by
\begin{eqnarray}
\mathcal{L}_0 
&=& 
\bar{q}(  i\slashed\partial - m_0)q
\\
\mathcal{L}_{q\bar{q}}
&=&
G_S\left[\left(\bar{q}q\right)^2
+\left(\bar{q}i\gamma_5{\bf \tau}q\right)^2\right]
\\
\mathcal{L}_{qq}
&=&
G_D\big\{\bar{q}\left[ i\gamma_5 C \tau_2 \lambda_2 \right]\bar{q}^T \big\}
        \left\{ q^T \left[ iC\gamma_5 \tau_2 \lambda_2 \right] q \right\},
\end{eqnarray}
where $\gamma_\nu$ are the Dirac matrices, $\tau_i$ are $SU(2)$ flavor matrices
and $C = i\gamma^2\gamma^0$ is the charge conjugation matrix.
$G_S$ and $G_D$ are the coupling strengths corresponding to 
the different channels, see Ref.~\cite{Buballa:2003qv} for a recent review. 
For the numerical analysis we adopt parameters from Ref. 
\cite{Grigorian:2006qe} and consider $\eta_D=G_D/G_S$  as a free parameter of the model.

A general method to deal with four-fermion interactions in the path integral 
approach starts with the Hubbard-Stratonovich transformation  
\cite{Kleinert:1977tv} of the 
partition function to its equivalent form in terms of collective bosonic 
fields, which is more suitable to deal with nonperturbative effects such as the
occurrence of order parameters related to phase transitions in the system 
as well as collective excitations (plasmons = mesons and pairs = diquarks)
in these phases.  

\subsection{Hubbard-Stratonovich transformation: Bosonization}
{
The Hubbard-Stratonovich transformation is a two-step procedure which 
consists of 
(1) linearization of the four-fermion interaction terms by 
introducing bosonic auxiliary fields in the appropriate channels and 
(2) integrating out the fermions analytically. 

We introduce the Hubbard-Stratonovich auxiliary fields 
$\Delta(\tau,x)$, $\Delta^*(\tau,x)$, $\pi(\tau, x)$ and $\sigma(\tau,x)$ so 
that the partition function of the system becomes
\begin{eqnarray}
\mathcal{Z}
&=&
\int\mathcal{D}\Delta^*\mathcal{D}\Delta\mathcal{D}\sigma\mathcal{D}\pi
\bigg\{e^{-\int^\beta d^{4}x\left[
        \frac{\sigma^2 + \pi^2}{4G_S} + \frac{|\Delta|^2}{4G_D}
\right]}
\\\nonumber
&&\times
\int\left[dq\right]\left[d\bar{q}\right]
e^{\int^\beta d^4x
        \left(\bar{q} (i\slashed\partial + \mu\gamma_0 - m_0)q
        - {\bar q}(\sigma+i\gamma_5{\bf \tau}\cdot{\bf \pi})q
        - \frac{\Delta^*}{2}q^T R q
        - \frac{\Delta}{2}\bar{q} \tilde{R} \bar{q}^T\right)}
\bigg\}\,.
\end{eqnarray}
where $R=iC\gamma_5\otimes \tau_2\otimes \lambda_2$, 
$\tilde{R}=i\gamma_5C\otimes\tau_2\otimes \lambda_2$.
By introducing Nambu-Gorkov spinors
\begin{eqnarray}
\Psi\equiv\frac{1}{\sqrt{2}}\left(
\begin{array}{c}
        q \\q^{c}
\end{array}
\right)
~~ , ~~
\bar\Psi\equiv
\frac{1}{\sqrt{2}}
\left(
\begin{array}{cc}
        \bar q \bar{q}^c
\end{array}
\right)
\end{eqnarray}
with $q^c (x)\equiv C \bar{q}^T (x)$, the Lagrangian takes the bilinear form
\begin{eqnarray}
\mathcal{L}=\bar\Psi\,\left(
\begin{array}{cc}
        i\slashed\partial + \mu\gamma_0 - \hat m - i\gamma_5{\bf \tau}\cdot{\bf \pi} 
        & \hspace{-1cm}
        i\Delta\gamma_5\tau_2\lambda_2 \\
        \hspace{-1cm} i\Delta^*\gamma_5\tau_2\lambda_2 
        &\hspace{-1cm}
        i\slashed\partial - \mu\gamma_0 - \hat m - i\gamma_5{\bf \tau}\cdot{\bf \pi} 
\end{array}
\right)
\Psi\,
\end{eqnarray}
with $\hat m = m_0 + \sigma$.
Hence the partition function becomes a Gaussian path integral in the 
bispinor fields which can be evaluated and yields the fermion determinant
\begin{eqnarray}
\mathcal{Z}
&=&
\int\mathcal{D}\Delta^*\mathcal{D}\Delta\mathcal{D}\sigma\mathcal{D}\pi
e^{-\int^{\beta}d^{4}x\frac{\sigma^2 
+ \pi^2}{4G_S}+\frac{|\Delta|^2}{4G_D}}
\int\mathcal{D}\bar\Psi\mathcal{D}\Psi
e^{\int^\beta d^{4}x\bar{\Psi}\left[S^{-1}\right]\Psi}
\\
&=&
\int\mathcal{D}\Delta^*\mathcal{D}\Delta\mathcal{D}\sigma\mathcal{D}\pi
e^{-\int^\beta d^4x\frac{\sigma^2 + \pi^2}{4G_S}
+\frac{|\Delta|^2}{4G_D}}
\cdot
{\rm Det}[S^{-1}]~,
\label{detglei}
\end{eqnarray}
where the inverse bispinor propagator is a matrix in Nambu-Gorkov-, Dirac-, 
color- and flavor space which after Fourier transformation reads
\begin{eqnarray}
\label{inv-prop}
S^{-1}
&=&
\left(
\begin{array}{cc}
(i\omega_n+\mu)\gamma_0 - \hat m-i{\bf \gamma}{\bf p} 
-i\gamma_5{\bf \tau}\cdot{\bf \pi}
& \hspace{-1.5cm} i\Delta\gamma_5\tau_2\lambda_2 \\
 \hspace{-1.5cm} i\Delta^*\gamma_5\tau_2\lambda_2 
& \hspace{-2cm} (i\omega_n-\mu)\gamma_0 
- \hat m-i{\bf \gamma}{\bf p} +i\gamma_5{\bf \tau}\cdot{\bf \pi}
\end{array}
\right)~.
\end{eqnarray}
So far we could derive with (\ref{detglei}) a very compact, bosonized form of 
the quark matter partition function (\ref{LL}) which is an exact 
transformation of (\ref{LL}), now formulated in terms of collective, 
bosonic fields.
As we will demonstrate in the following, this form is suitable since it allows
to obtain nonperturbative results already in the lowest orders with respect to
an expansion around the stationary values of these fields. In performing this 
expansion, we may factorize the partition function into mean field (MF),
Gaussian fluctuation (Gauss) and residual (res) contributions
$$
  Z(\mu,T)\equiv e^{-\beta\Omega(\mu,T)}
=Z_{MF}(\mu,T) Z_{\rm Gauss}(\mu,T)Z_{\rm res}(\mu,T)~.
$$
In the following we will discuss the physical content of these approximations.
}
\subsection{Mean-field approximation: order parameters}{
In thermodynamical equilibrium, the mean field values satisfy the 
stationarity condition of the minimal thermodynamical potential 
$\Omega_{MF}\equiv-\frac{1}{\beta V}\ln\mathcal{Z}_{MF}$, i.e.
\begin{equation}
\frac{\partial\Omega_{MF}}{\partial\sigma_{MF}}=
\frac{\partial\Omega_{MF}}{\partial\pi_{MF}}=
\frac{\partial\Omega_{MF}}{\partial\Delta_{MF}}=0~,
\label{Ex-OmegaMF}
\end{equation}
equivalent to the fulfillment of the gap equations 
$\sigma_{MF}=-4G_S{\rm Tr}\left(S_{MF}\right)\equiv m-m_0$, 
${\vec{\pi}}_{MF}=-4iG_S{\rm Tr}\left(\gamma_5{\vec{\tau}}S_{MF}\right)
=0$ 
and 
$\Delta_{MF}=4G_D{\rm Tr}\left(\gamma_5\tau_2 \lambda_2 S_{MF}\right)
=\Delta$, 
together with the stability criterion that the determinant of the 
curvature matrix formed by the second derivatives is positive.
After the evaluation of the traces in the internal spaces and the sum over 
the Matsubara frequencies one gets
\begin{eqnarray}
\label{OmegaMF}
\Omega_{MF} 
&=& 
-\frac{1}{\beta V}\ln\mathcal{Z}_{MF}
=
\frac{(m - m_0)^2}{4G_S} + \frac{|\Delta|^2}{4G_D} 
- \frac{1}{\beta V}{\rm Tr}\left(\ln\beta S_{MF}^{-1}\right)
\nonumber\\
&=&\hspace{-2mm}
\frac{(m - m_0)^2}{4G_S} + \frac{|\Delta|^2}{4G_D} -4\int\frac
{d^3p}{(2\pi)^3}\left[
E_{\bf p}^+ + E_{\bf p}^- + E_{\bf p} + 2T\ln(1+e^{-\beta E_{\bf p}^+})
\right.
\nonumber\\
&&\hspace{-2mm}
+ \left. 2T\ln(1+e^{-\beta E_{\bf p}^-}) 
+ T\ln(1+e^{-\beta \xi_{\bf p}^+})
+ T\ln(1+e^{-\beta \xi_{\bf p}^-})\right]
\label{thermodynamic-potential-MF}
\end{eqnarray}
where we have defined the particle dispersion relation 
$E_{\bf p}^\pm = \sqrt{\left(\xi_{\bf p}^\pm \right)^2 + \Delta^2}$ with 
$\xi_{\bf p}^\pm = E_{\bf p}\pm\mu$, $E_{\bf p} = \sqrt{m^2+{\bf p}^2}$. 
The $\Delta\neq 0$ dispersion law is associated to the red and green quarks 
$(E_{\bf p}^{-})$ and antiquarks $(E_{\bf p}^+)$, whereas the ungapped 
blue quarks (antiquarks) have the dispersion 
$\xi^-_{\bf p}$ ($(\xi^+_{\bf p})$).
From Eqs. (\ref{Ex-OmegaMF}) with (\ref{OmegaMF}) we obtain the gap 
equations for the order parameters $m$ and $\Delta$, which have to be 
solved self-consistently,
\begin{eqnarray}
m-m_0 &=& 8G_{S}\,m\int\frac{d^3p}{(2\pi)^3}\frac{1}{E_{\bf p}}
\bigg\{
\left[ 1 - 2n_F(E_{\bf p}^-) \right]\frac{\xi_{\bf p}^-}{E_{\bf p}^-} 
\nonumber \\ && \hspace{1.5cm}
+\left[ 1 - 2n_F(E_{\bf p}^+) \right]\frac{\xi_{\bf p}^+}{E_{\bf p}^+}
+ n_F(-\xi_{\bf p}^+) -  n_F(\xi_{\bf p}^-)\bigg\}~,
\\
\Delta &=& 8G_{D} \int\frac{d^3p}{(2\pi)^3}\left[
\frac{1 - 2n_F(E_{\bf p}^-)}{E_{\bf p}^-} + \frac{1 - 2n_F(E_{\bf p}^+)}{E_{\bf p}^+}
\right]~,
\end{eqnarray}
with the Fermi dsitribution function $n_F(E) = (1+e^{\beta E})^{-1}$. 
For zero temperature, the gap equations take the simple form
\begin{eqnarray}
\label{eq:m-m0}
m-m_0 
&=&
8G_{S} m\int\frac{d^3p}{(2\pi)^3}\frac{1}{E_{\bf p}}
\left[\frac{\xi_{\bf p}^-}{E_{\bf p}^-}+\frac{\xi_{\bf p}^+}{E_{\bf p}^+} 
+ \Theta(\xi_{\bf p}^-)\right]~,
\\
\Delta &=& 8G_{D}\,\Delta\int\frac{d^3p}{(2\pi)^3}
\left[\frac{1}{E_{\bf p}^-} + \frac{1}{E_{\bf p}^+} \right]\,.
\label{eq:Delta_MF}
\end{eqnarray}
Solutions of the gap equations for the dynamically generated quark mass $m$
and for the diquark pairing gap $\Delta$ at 
$T = 0$ as a function of the chemical potential
are shown in Fig.~\ref{order}.
\begin{figure}
\epsfig{figure=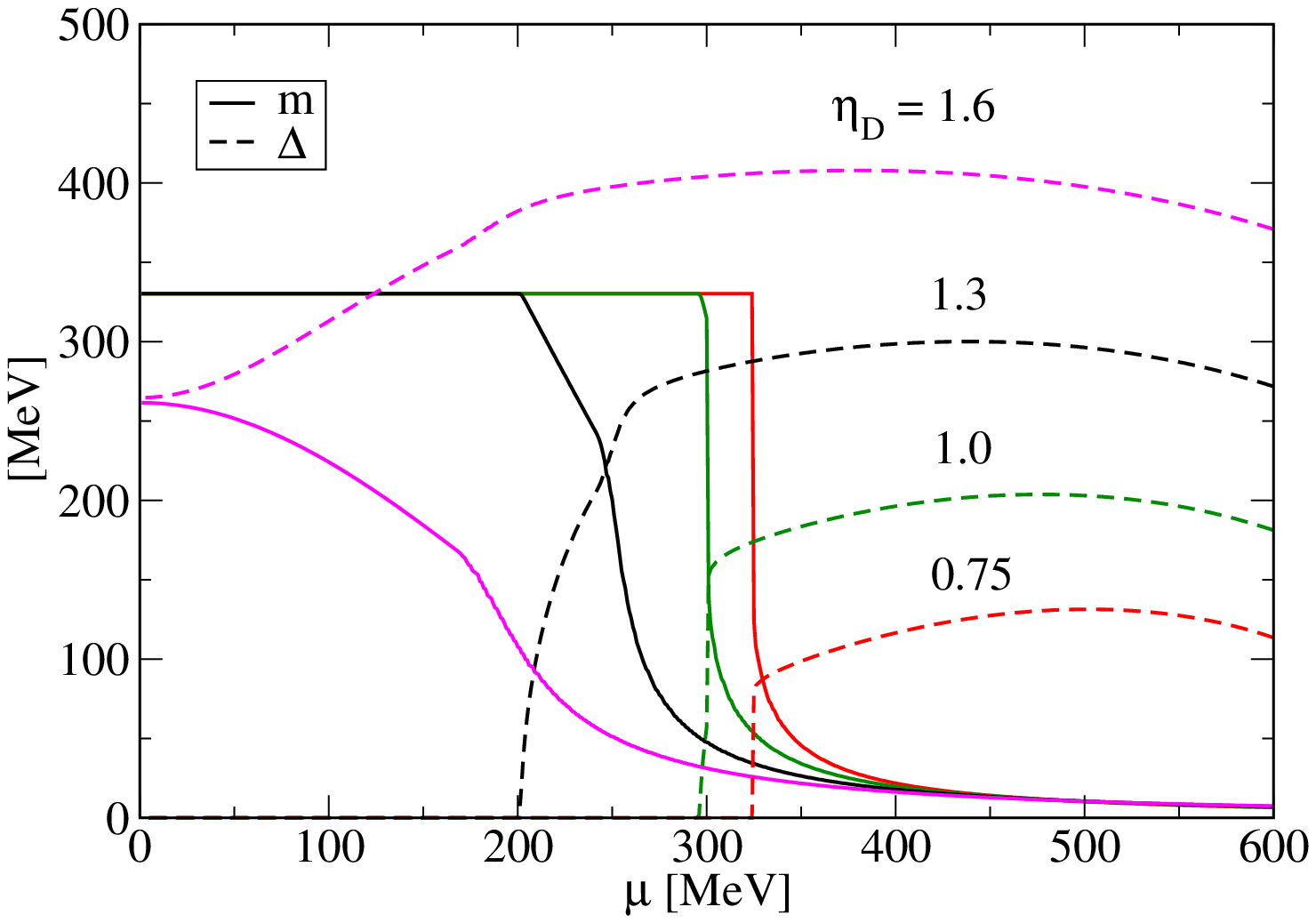,height=0.75\textwidth,width=0.5\textwidth,angle=-90}
\caption{Order parameters for chiral symmetry breaking (full lines) and 
color superconductivity (dashed lines) at $T = 0$ for 
different values of the diquark coupling $\eta_D$. First order phase 
transitions turn to second order or even crossover when $\eta_D$ is increased.
For details, see text.
\label{order}}
\end{figure}
\begin{figure}
\epsfig{figure=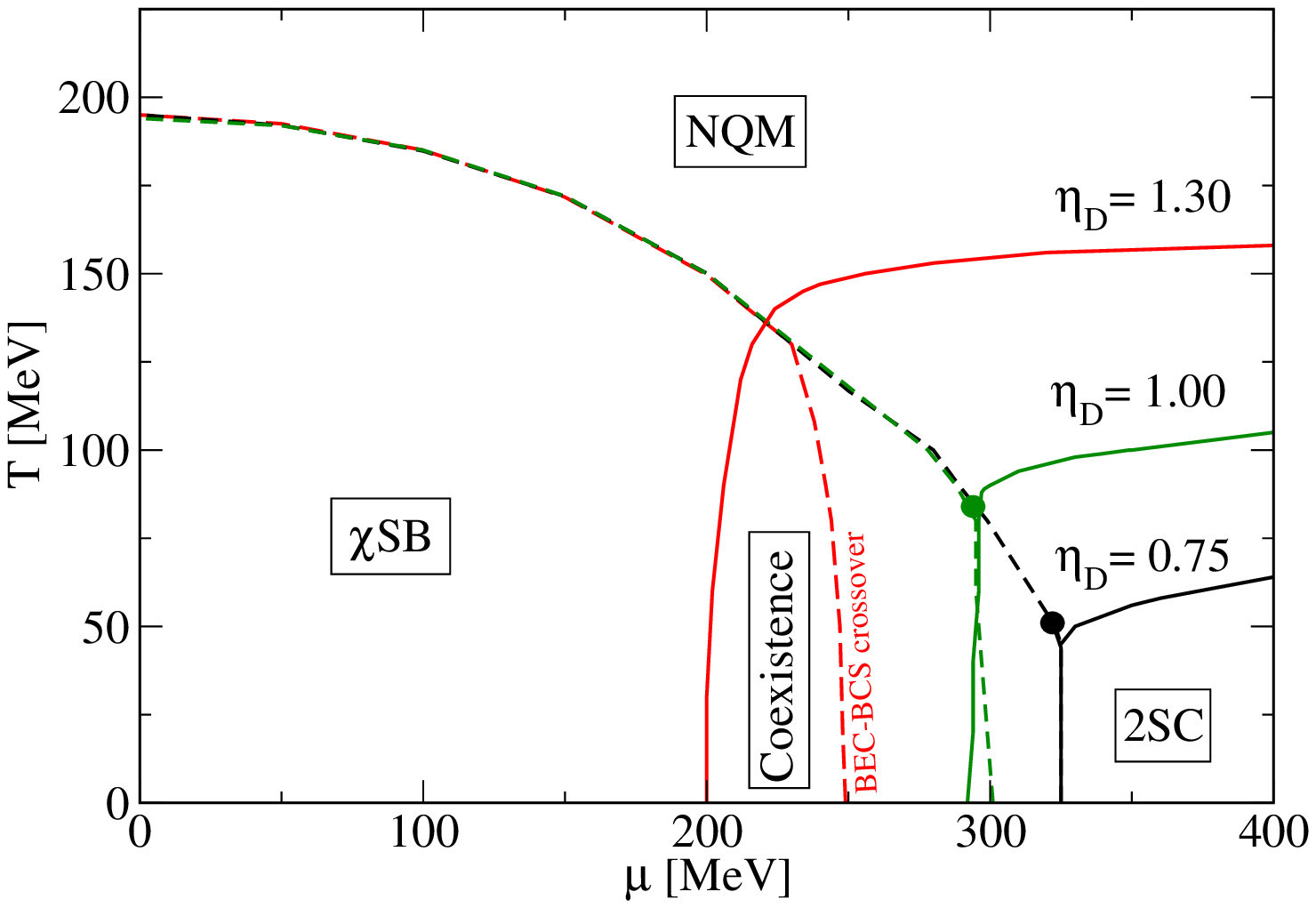,height=0.75\textwidth,width=0.55\textwidth,angle=-90}
\caption{Phase diagram of two-flavor quark matter with critical lines 
for chiral symmetry breaking (dashed) and color superconductivity (solid)
for three values of the diquark coupling strength: $\eta_D=0.75$ (black),  
$1.0$ (green) and $1.3$ (red). 
The BEC-BCS crossover occurs when the chiral transition 
(coincident with the Mott transition for mesonic and diquark bound states) 
occurs inside the 2SC phase. It is characterized 
by the coexistence of diquark condensation with chiral symmetry breaking. 
}
\label{phases}
\end{figure}
From the knowledge of the order parameters as functions of the 
thermodynamical variables ($T, \mu$) we can deduce the phase diagram
of  Fig.~\ref{phases}.
}
\subsection{Phase diagram}{
From the solutions of the gap equations for the order parameters 
in dependence of the thermodynamical variables $T$ and $\mu$ we have 
constructed the phase diagram of the present quark matter model in the 
$T-\mu$ plane, see Fig.~\ref{phases}.
The two order parameters allow to distinguish 4 phases:\pagebreak
\begin{itemize}
\item $\Delta=0$, $m\sim m_0$: normal phase (NQM)
\item $\Delta\neq 0$, $m\sim m_0$: color superconductor (2SC)
\item $\Delta=0$, $m \gg m_0$: chiral symmetry broken phase ($\chi$SB)
\item $\Delta\neq 0$, $m\gg m_0$: coexistence of $\chi$SB and 2SC 
(BEC phase)
\end{itemize}
Order parameters are indicators of phase transitions. 
The phase transitions can be classified according to their order,
depending on the behavior of the order parameters with the change of
thermodynamic variables
\begin{itemize}
\item first order: order parameter jumps, as in the case of the 
$\chi$SB $\to$ 2SC phase transition at not too large coupling.
\item second order: order parameter turns continuously to zero.
The 2SC $\to$ NQM transition with increasing $T$ is always second order. 
The $\chi$SB $\to$  2SC transition turns from 1st to 2nd order for strong 
enough coupling. 
\item crossover: the order parameter changes continuously, but does not 
go to zero and also has no jumps. An example is the $\chi$SB $\to$ NQM 
transition at temperatures above the critical endpoint (CP) where it goes over
to the line of 1st order transitions in the $T-\mu$ plane.
The identification of the CP is a key issue for experimental research and 
thus for the verification of QCD models.
It is suggested that verifiable signatures (change of the fluctuation spectrum,
latent heat or not) are related with it.
For strong coupling, the CP moves to lower $T$ and finally to $T = 0$
(for $\eta_D > 1.3$ the chiral transition is always crossover).
\end{itemize}

Increasing the diquark coupling $\eta_D$ leads to an increase of the 
diquark gap and therefore a rise in the critical temperature for the 
second order transition to a normal quark matter phase. 
It shifts also the border between color superconductivity (2SC) and chiral 
symmetry broken phase ($\chi$SB) to lower values of the chemical potential. 
For very strong coupling $\eta_D\sim 1$, a coexistence region developes, 
where both order parameters are simultaneously nonvanishing. 
Under these conditions, the phase border is not of first order and therefore
no critical endpoint can be identified. 
As we are going to explain in the next section, in the  $\chi$SB phase pion 
and diquark bound states can exist. At the chiral symmetry restoration 
transition, they merge the continuum of unbound states and turn into (resonant)
scattering states. 
When this Mott transition occurs within the 2SC phase (characterized by a 
nonvanishing diquark condensate) we speak of a BEC-BCS crossover:
the condensation of diquark bound states (BEC) turns into a condensation of 
resonances, called Cooper pairs (BCS).

For lower coupling, the critical point occurs and is shown as a colored dot 
in the phase diagram of Fig.~\ref{phases}.

In the next section we will turn towards the interesting question about the 
quasiparticle excitations in these phases.  
To this end, we will expand the action functional in the partition function 
up to quadratic (Gaussian) order in the mesonic fields and arrive at a 
tractable approximation for the bosonized quark matter model 
(\ref{detglei}).
}
\subsection{Gaussian fluctuations: bound \& scattering states}{
Let us expand now the mesonic fields around their mean field values.  
In these lectures, we will focus on fluctuations in the mesonic channels,
where the pion and the sigma meson will emerge as quasiparticle degrees of 
freedom. 
On the example of the pion we will explain the physics of the Mott transition.
As discussed in the previous section, the phenomenon of the BEC-BCS crossover 
in the 2SC phase is due to the Mott transition for diquarks.
The detailed investigation of the quantized diquark fluctuations, which are 
also a prerequisite of the formation of baryons, will be given elsewhere 
\cite{Sun:2007fc,Ebert:2004dr}.
As we already noticed, the pion does not contribute to the mean field, and 
we need to introduce only the sigma-field fluctuations as
$\sigma\rightarrow\sigma_{MF}+\sigma$.  
Hence it is possible to decompose the inverse 
propagator $S^{-1}$ into a mean field part and a fluctuation part 
$S^{-1}=S^{-1}_{MF}+\Sigma\,$, where the matrix $\Sigma$ is defined as 
\begin{eqnarray}
\Sigma\equiv\left(
\begin{array}{cc} 
-\sigma - i\gamma_5{\vec{\tau}}\cdot{\vec{\pi}} & 0 \\ 
0  & -\sigma - i\gamma_5{\vec{\tau}}^{t}\cdot{\vec{\pi}}
\end{array}
\right)~.
\end{eqnarray}
In the Gaussian approximation the fermion determinant becomes
\begin{eqnarray}
\frac{{\rm Det} \left[S^{-1}\right]\big|_{\rm Gauss}}
{{\rm Det} \left[S^{-1}_{MF}\right]}
=
\exp\left\{-\frac{1}{2}\int\frac{d^{4}q}{(2\pi)^4}
{\rm Tr}\left[S_{MF}(p)\Sigma(q) S_{MF}(p+q)\Sigma(q)\right]\right\}~.
\label{expansion-propagator}
\end{eqnarray}
The propagator $S_{MF}$ is obtained from (\ref{inv-prop}) by the matrix 
inversion\footnote{contribution by M. Buballa}
\begin{equation}
S_{MF}\equiv
\left(\begin{array}{cc}
{\bf G}^+ & {\bf F}^- \\
{\bf F}^+ & {\bf G}^-
\end{array}\right)
\end{equation}
with the matrix elements
\begin{eqnarray}
{\bf G}^{\pm}_{p}
&=&
\sum_{s_p}\sum_{t_p}\frac{t_p}{2E_{\bf p}^{\pm s_p}}
\frac{t_pE_{\bf p}^{\pm s_p}-s_p\xi_{\bf p}^{\pm s_p}}
       {p_0-t_pE_{\bf p}^{\pm s_p}}
\Lambda_{\bf p}^{-s_p}\gamma_0\mathcal{P}_{\rm rg}
+\sum_{s_p}
\frac{\Lambda_{\bf p}^{-s_p}\gamma_0\mathcal{P}_{\rm b}}
       {p_{0}+s_p\xi_{\bf p}^{\pm s_p}} \,,\\
{\bf F}^\pm_p
&=&
i\sum_{s_p}\sum_{t_p}
\frac{t_p}{2 E_{\bf p}^{\pm s_p}}
\frac{\Delta^\pm}{p_0-t_pE_{\bf p}^{\pm s_p}}
\Lambda_{\bf p}^{s_p}\,\gamma_5\tau_2\lambda_2\,,
\end{eqnarray}
where $s_p,t_p=\pm1$, $(\Delta^+,\Delta^-)=(\Delta^*,\Delta)$.
For the subsequent evaluation of traces in quark-loop diagrams, 
it is convenient to use this notation with projectors in color space, 
$\mathcal{P}_{\rm rg}={\rm diag}(1,1,0)$, 
$\mathcal{P}_{\rm b}={\rm diag}(0,0,1)$ and in Dirac space,
$$\Lambda^\pm_{\bf p}=
\frac{1}{2}\left[1\pm\gamma_0
\left(\frac{{\vec \gamma}\cdot{\vec p}+\hat m}{E_{\bf p}}\right)\right]\,.$$
The summation over  Matsubara frequencies $p_0=i\omega_n$ is most 
systematic using the above decomposition into simple poles in the $p_0$ plane.
The poles of the normal propagators $\bf G^\pm$  are given by the gapped
dispersion relations for the paired  red-green quarks (antiquarks), 
$E^\pm_{\bf p} = \sqrt{(\xi^\pm_{\bf p})^2+\Delta^2}$, 
and the ungapped dispersions $\xi^{\pm}_{\bf p} = E_{\bf p}\pm \mu$ for the 
blue quarks (antiquarks). 
The anomalous propagators $\bf F^\pm_{\bf p}$ are only nonvanishing in the 2SC phase
when the pair amplitude is nonvanishing.
Let us notice explicitly that this procedure has yielded an effective action 
that includes the fluctuation terms responsible for the excitation of 
scalar and pseudoscalar mesonic modes. 
The evaluation of the traces (\ref{expansion-propagator}) can be performed 
with the result
\begin{eqnarray}
\frac{1}{2}{\rm Tr}\left(S_{MF}\Sigma S_{MF}\Sigma\right)=
\left(
{\vec \pi}, \sigma
\right)
\left(
\begin{array}{cc}
\Pi_{\pi\pi}&0\\
0&\Pi_{\sigma\sigma}\\
\end{array}
\right)
\left(
\begin{array}{c}
{\vec \pi}\\ \sigma
\end{array}
\right)
\end{eqnarray}
with 
\begin{eqnarray}
\Pi_{\sigma\sigma}(q_0,{\bf q})
&\equiv&
{\rm Tr}[
        {\bf G}^+_p{\bf G}^+_{p+q}
        +{\bf F}^-_p{\bf F}^+_{p+q}
        +{\bf G}^-_p{\bf G}^-_{p+q}
        +{\bf F}^+_p{\bf F}^-_{p+q}
]
\\\nonumber
\Pi_{\pi\pi}(q_0,{\bf q})
&\equiv&
-{\rm Tr}[{\bf G}^+_p(\gamma_5{\vec\tau}){\bf G}^+_{p+q}
(\gamma_5{\vec \tau})
+{\bf F}^-_{\bf p}(\gamma_5{\vec\tau}^t)
{\bf F}^+_{p+q}(\gamma_5{\vec \tau})\\
&&
+{\bf F}^+_p(\gamma_5{\vec \tau})
{\bf F}^-_{p+q}(\gamma_5{\vec \tau}^t)
+{\bf G}^-_p(\gamma_5{\vec \tau}^t)
{\bf G}^-_{p+q}(\gamma_5{\vec \tau}^t)]~.
\end{eqnarray}
These polarization functions are the key quantities for the 
investigation of mesonic bound and scattering states in quark matter.
In the following we perform the further evaluation and discussion for the
pionic modes, the $\sigma$ modes is treated in an analogous way.
We start with the evaluation of traces and Matsubara summation.
\begin{eqnarray} 
&& 
\Pi_{\pi\pi}(q_0,{\bf q}) 
= 
2\int\frac{d^3p}{(2\pi)^3}\sum_{s_p,s_k} 
\mathcal{T}_-^+ (s_p,s_k) \bigg\{ 
        \frac{n_F(s_p\xi_{\bf p}^{s_p}) - n_F(s_k\xi_{\bf p+q}^{s_k})} 
               {q_0 - s_k\xi_{\bf p+q}^{s_k} + s_p\xi_{\bf p}^{s_p}}  
\nonumber \\  
&&\hspace{0.5cm}  
+ \frac{n_F(s_p\xi_{\bf p}^{s_p})-n_F(s_k\xi_{\bf p+q}^{s_k})} 
               {q_0 + s_k\xi_{\bf p+q}^{s_k} - s_p\xi_{\bf p}^{s_p}} 
   +  
        \sum_{t_p,t_k} 
        \frac{t_pt_k}{E_{\bf p}^{s_p}E_{\bf p+q}^{s_k}} 
        \frac{n_F(t_pE_{\bf p}^{s_p})-n_F(t_kE_{\bf p+q}^{s_k})} 
               {q_0 - t_kE_{\bf p+q}^{s_k} + t_pE_{\bf p}^{s_p}}     
\nonumber \\ 
&& \hspace{0.5cm}
\times \left(t_pt_kE_{\bf p}^{s_p}E_{\bf p+q}^{s_k} 
+ s_ps_k\xi_{\bf p}^{s_p}\xi_{\bf p+q}^{s_k} - |\Delta|^2 \right) 
\bigg\} 
\end{eqnarray} 
where
\begin{eqnarray}
\mathcal{T}_-^+ (s_p,s_k)
=
\left(1+ s_ps_k\frac{{\bf p}\cdot({\bf p+q})-m^2}{E_{\bf p}E_{\bf p+q}}\right)\,.
\end{eqnarray}
For a pionic mode at rest in the medium (${\bf q}=0$) this reduces to
\begin{eqnarray} 
&& \Pi_{\pi\pi}(q_0,{\bf 0}) =
8\int\frac{d^3p}{(2\pi)^3} 
\bigg\{ 
    N(\xi_{\bf p}^+,\xi_{\bf p}^-)
    \left[\frac{1}{q_0 - 2E_{\bf p}} - \frac{1}{q_0 + 2E_{\bf p}}\right] 
\nonumber \\
&& \hspace{-5mm} + 
\left[1- \frac{\xi_{\bf p}^+\xi_{\bf p}^- + \Delta^2}
{E_{\bf p}^+E_{\bf p}^-}\right] 
    M(E_{\bf p}^+, E_{\bf p}^-) 
   \left[\frac{1}{q_0 - E_{\bf p}^+ + E_{\bf p}^-} 
 - \frac{1}{q_0 + E_{\bf p}^+ - E_{\bf p}^-}\right]  
\nonumber \\
&& \hspace{-5mm} + 
 \left[1+\frac{\xi_{\bf p}^+\xi_{\bf p}^- +\Delta^2}{E_{\bf p}^+E_{\bf p}^-}
\right]
 N(E_{\bf p}^+, E_{\bf p}^-)
 \left[\frac{1}{q_0 - E_{\bf p}^+ - E_{\bf p}^-} 
- \frac{1}{q_0 + E_{\bf p}^+ + E_{\bf p}^-}\right] 
\bigg\}
\label{mu1}
\end{eqnarray} 
where we have introduced the phase space occupation factors 
$N(x,y) = 1-n_F(x)-n_F(y)$ (Pauli blocking) and $M(x,y) = n_F(x)-n_F(y)$.
For $\mu \neq 0$ this function has three poles from the first terms in each 
bracket, corresponding to positive energies ($q_0 > 0$).  
So we need to focus only on these three terms. 
For $\mu = 0$ the second term vanishes due to the prefactor and we are 
left with two poles.

We make use of the Dirac identity 
$\lim_{\eta\to 0}\frac{1}{x + i\eta} = \mathcal{P}\frac{1}{x} - i\pi\delta(x)$ 
in order to decompose the polarization function into real and imaginary 
parts after analytical continuation to the complex plane.
The imaginary part is starightforwardly integrated after transformation from
momentum to energy $\omega$.
At the pole the variables transform as 
\begin{eqnarray} 
p_\omega  
&=& 
\sqrt{\frac{\omega^4 - 4\omega^2(\mu^2 + \Delta^2)}
               {4(\omega^2 - 4\mu^2)} - m^2 }~.
\end{eqnarray} 
For $\eta_D < 1$ we know that $\Delta = 0$ if $m \geq \mu$, what includes 
that $\omega \geq 2\mu$ as this is the relevant threshold. 
Therefore, the pole is not hidden and we recover the usual $2m$ threshold. 
For small enough couplings, $\Delta \neq 0$ only if $m < \mu$. 
Therefore, this pole is not hidden in this case. 
This reasoning includes that the argument of the square root is strictly 
positive. 
The integration borders thus shift 
$p\in (0,\infty) \to \omega\in (X_\pm,\infty)$, where 
the thresholds are given by $2m$ and
\begin{eqnarray} 
X_\pm 
&=& 
        \sqrt{(m+\mu)^2+\Delta^2}\pm\sqrt{(m-\mu)^2+\Delta^2} ~.
\label{mu2}
\end{eqnarray} 
The pion polarization function in the 2SC phase can thus be decomposed into
real and imaginary parts in the following form
\begin{eqnarray} 
&&\Pi_{\pi\pi}^\Delta(\omega+i\eta, {\bf 0})
= 
{\bf Re}\Pi_{\pi\pi}^\Delta(\omega+i\eta, {\bf 0}) 
+ i{\bf Im}\Pi_{\pi\pi}^\Delta(\omega+i\eta, {\bf 0})
\nonumber \\
&& = 
8\int\frac{d^3p}{(2\pi)^3} 
\bigg\{ 
    N(\xi_{\bf p}^+, \xi_{\bf p}^-)
    \left[
        \frac{\mathcal{P}}{\omega - 2E_{\bf p}} 
        -
        \frac{1}{\omega + 2E_{\bf p}}
    \right] 
\nonumber \\
&& + 
    \left[
        1-\frac{\xi_{\bf p}^+\xi_{\bf p}^- + \Delta^2}{E_{\bf p}^+E_{\bf p}^-}
    \right]
    M(E_{\bf p}^+, E_{\bf p}^-) 
    \left[
        \frac{\mathcal{P}}{\omega - E_{\bf p}^+ + E_{\bf p}^-} 
        -
        \frac{1}{\omega + E_{\bf p}^+ - E_{\bf p}^-}
    \right] 
\nonumber \\
&& + 
    \left[
        1+\frac{\xi_{\bf p}^+\xi_{\bf p}^- + \Delta^2}{E_{\bf p}^+E_{\bf p}^-} 
    \right]
    N(E_{\bf p}^+, E_{\bf p}^-)
    \left[
        \frac{\mathcal{P}}{\omega - E_{\bf p}^+ - E_{\bf p}^-} 
        -
        \frac{1}{\omega + E_{\bf p}^+ + E_{\bf p}^-}
    \right] 
\bigg\} 
\nonumber \\ 
&& - 
i\frac{2}{\pi}
\bigg\{
p_\omega^0 E_{p_\omega^0}  
N(\xi_{p_\omega^0}^+, \xi_{p_\omega^0}^-)
\Theta(\omega - 2m) 
\nonumber \\
&& + 
p_\omega E_{p_\omega} 
 \frac{E_{p_\omega}^+E_{p_\omega}^- -\xi_{p_\omega}^+\xi_{p_\omega}^- -\Delta^2}
        {\xi_{p_\omega}^+ E_{p_\omega}^- - \xi_{p_\omega}^- E_{p_\omega}^+} 
    M(E_{p_\omega}^+, E_{p_\omega}^-)
    \Theta(\omega-X_-) \label{mu3} 
\nonumber \\
&& + 
p_\omega E_{p_\omega} 
    \frac{E_{p_\omega}^+E_{p_\omega}^- + \xi_{p_\omega}^+\xi_{p_\omega}^- + \Delta^2}
           {\xi_{p_\omega}^+ E_{p_\omega}^- + \xi_{p_\omega}^- E_{p_\omega}^+} 
    N(E_{p_\omega}^+, E_{p_\omega}^-)
    \Theta(\omega-X_+) 
\bigg\} 
\end{eqnarray} 
where $\mathcal{P}$ denotes the principal value integration, 
$p_\omega^0 = p_\omega\mid_{\Delta=0} = \sqrt{\frac{\omega^2}{4}-m^2}$ 
and we have made explicit the three thresholds, $2m$ and $X_\pm$, 
for the occurrence of the corresponding decay processes giving rise to 
the partial widths $\Gamma_{2m}$ and $\Gamma_\pm$, respectively. 
In the normal phase this reduces to 
\begin{eqnarray} 
\Pi_{\pi\pi}^0(\omega+i\eta, {\bf 0})
&=& 
{\bf Re}\Pi_{\pi\pi}^0(\omega+i\eta, {\bf 0})
+ i{\bf Im}\Pi_{\pi\pi}^0(\omega+i\eta, {\bf 0})
\nonumber \\
&=& 
24\int\frac{d^3p}{(2\pi)^3} 
    N(\xi_{\bf p}^+, \xi_{\bf p}^-)
    \left[\frac{\mathcal{P}}{\omega - 2E_{\bf p}} 
- \frac{1}{\omega + 2E_{\bf p}}\right] 
\nonumber \\
&&- 
i\frac{6}{\pi}p_\omega^0 E_{p_\omega^0} 
N(\xi_{p_\omega^0}^+, \xi_{p_\omega^0}^-)
\Theta(\omega-2m) 
\end{eqnarray} 
The analytic properties of the mesonic modes can be analyzed from their
spectral function. Here we discuss results for pionic modes with ${\bf q}=0$
in the rest frame of the medium
\begin{equation}
\rho_\pi(\omega+i\eta, {\bf 0}) 
=
\frac{8G^2_\sigma{\bf Im}\Pi_{\pi\pi}(\omega+i\eta, {\bf 0})}
{[1-2G_\sigma{\bf Re}\Pi_{\pi\pi}(\omega, {\bf 0})]^2
+[2G_\sigma {\bf Im}\Pi_{\pi\pi}(\omega+i\eta, {\bf 0})]^2}~.
\end{equation}
In the limit of vanishing imaginary part, we recover the spectral function
for a ``true'' (on-shell) bound state 
\begin{equation}
\lim_{{\bf Im}\Pi_{\pi\pi}\to 0}\rho_\pi(\omega+i\eta, {\bf 0})
=
2\pi\delta(1 - 2G_\sigma{\bf Re}\Pi_{\pi\pi}(\omega, {\bf 0}))~,
\end{equation}
which corresponds to an infinite lifetime of the state and a mass to be 
found from the pole condition 
$1-2G_\sigma{\bf Re}\Pi_{\pi\pi}(m_\pi, {\bf 0})=0$.
In Fig.~\ref{pion_mu000}, we show results for the mass spectrum of pions 
and sigma-mesons  as a function 
of the temperature for vanishing chemical potential $\mu_B = 0$ and strong 
diquark coupling $\eta_D = 1.0$. Since $\Delta = 0$, the only threshold for 
the imaginary parts of meson decays is  $2m$. 
The $\sigma$ mass is always above the threshold and therefore this state 
is unstable in the present model. The pion, however, is a bound state until 
the critical temperature for the Mott transition $T_{\rm Mott} = 212.7$ MeV
is reached. For $T > T_{\rm Mott}$ the pion becomes unstable for decay into
quark-antiquark pairs. As can be seen from the behavior of the spectral 
function in the lower panel of Fig.~\ref{pion_mu000}, the pion is still a
well-identifyable, long-lived resonance in that case. 
The detailed analytic behavior of the pion at the Mott transition has been
discussed in the context of the NJL model by H\"ufner et al. 
\cite{Hufner:1996pq}, see also the inset of the lower panel of 
Fig.~\ref{pion_mu000}. It shows strong similarities with the behavior of
bound states of fermionic atoms in traps when their coupling is tuned by
exploiting Feshbach resonances in an external magnetic field, see  
\cite{AnnPhys}.
In the context of RHIC experiments, one has discussed such quasi-bound 
states as an explanation for the perfect liquid behavior of the sQGP 
\cite{Shuryak:2004tx}.
\begin{figure}
\vspace{-1cm}
\epsfig{figure=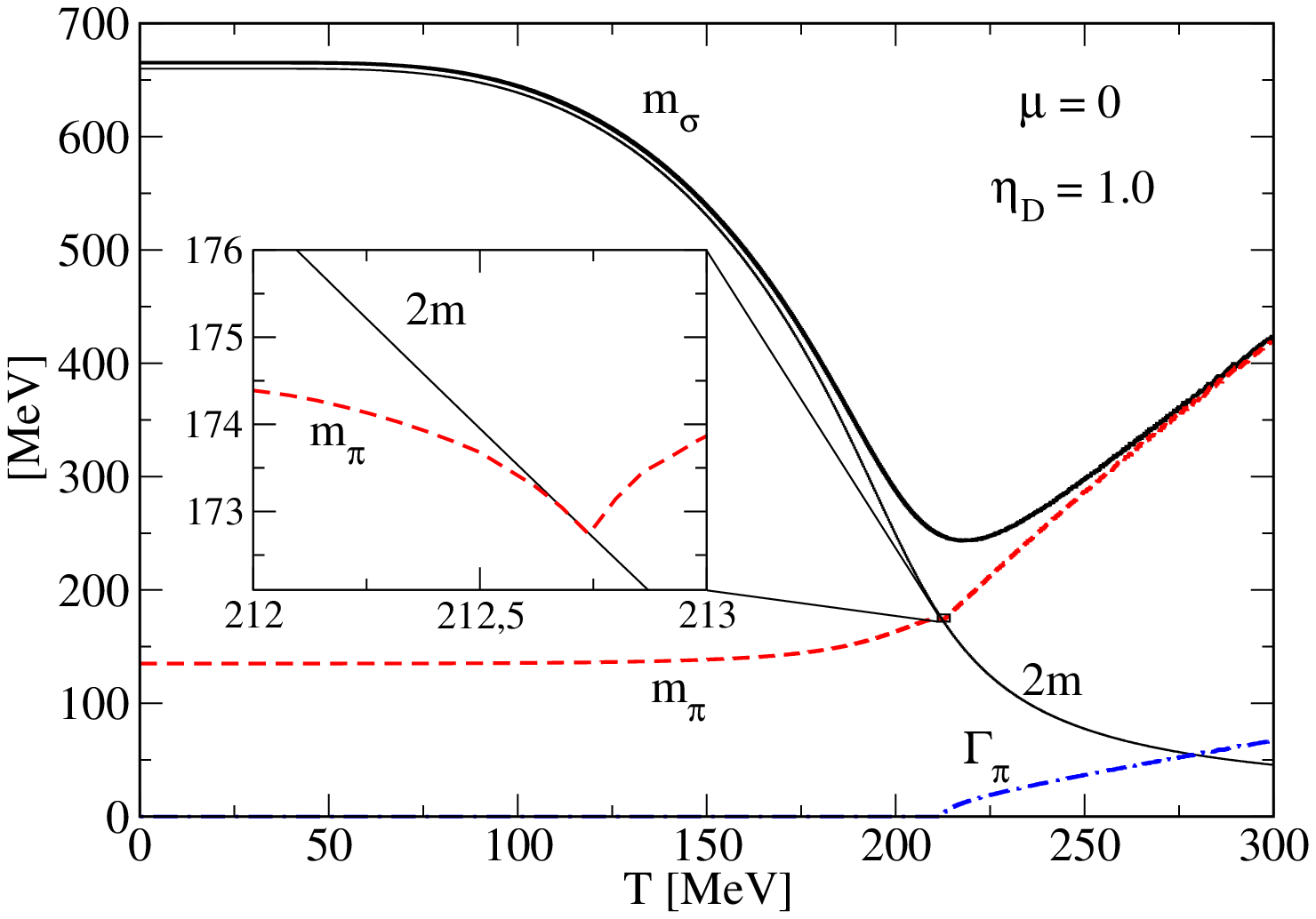,width=0.65\textwidth,angle=-90}
\vspace{-1cm}
\epsfig{figure=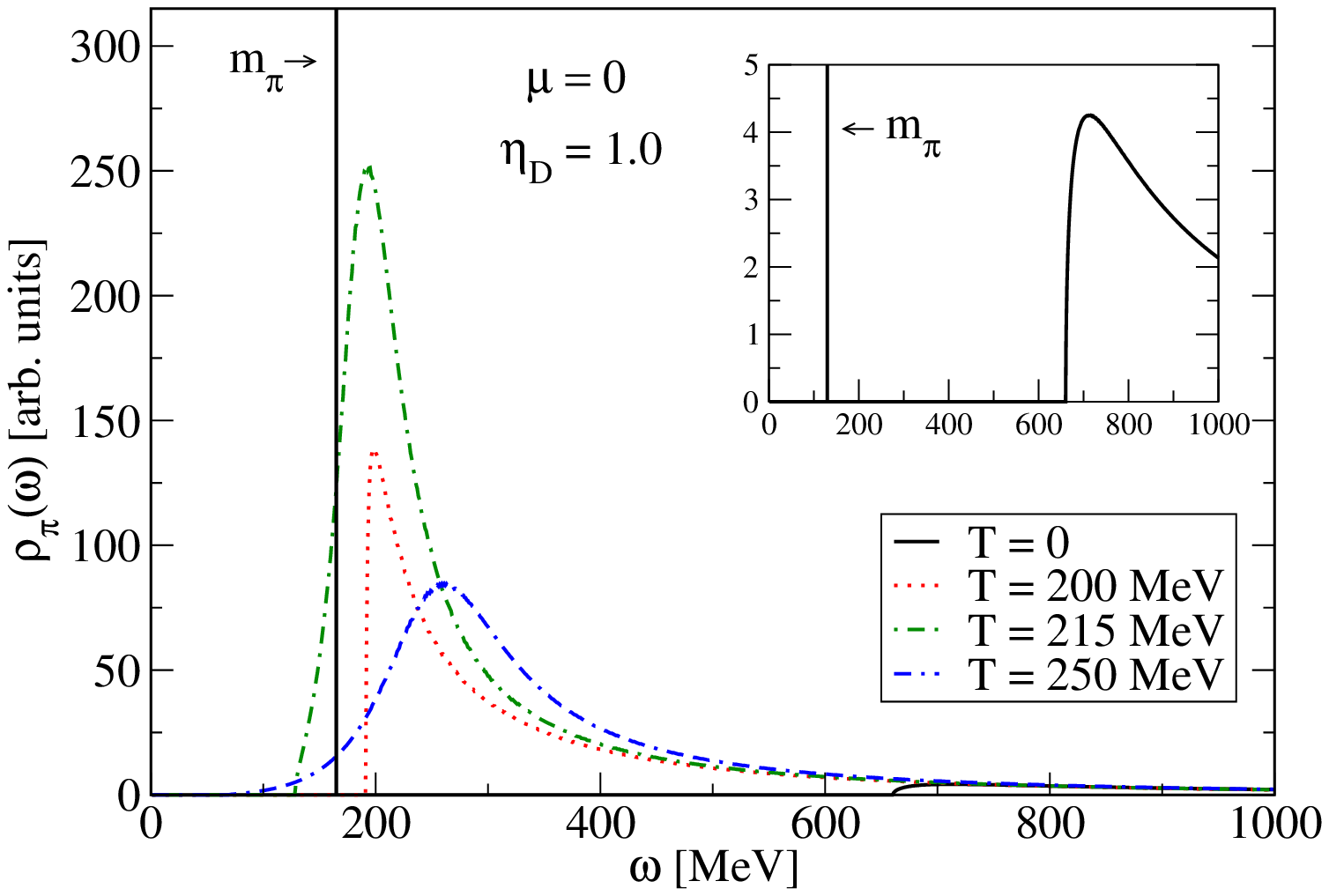,width=0.65\textwidth,angle=-90}
\caption{{\it Upper panel:} 
Mass spectrum of mesons ($\pi$, $\sigma$) as a function 
of the temperature for vanishing chemical potential $\mu_B = 0$ and strong 
diquark coupling $\eta_D=1.0$. 
The threshold  $E_{\rm th} = 2~m_q$ for Mott dissociation of pions and 
occurrence of a nonvanishing decay width $\Gamma_\pi=Im~\Pi_\pi/m_\pi$ 
is reached at $T_{\rm Mott} = 212.7$ MeV (see inset).
{\it Lower panel:}
Spectral function for pionic correlations for $\mu_B = 0$ in the vacuum
at $T = 0$ (see inset) and at different temperatures around the Mott transition.
Below $T_{\rm Mott}$, the bound state (delta function) and the
continuum of scattering states are separated by a mass gap. 
Above   $T_{\rm Mott}$, the spectral function is still sharply peaked, 
related to a lifetime of pionic correlations of the order of the lifetime
of a fireball in heavy-ion collisions (quasi-bound states in the quark plasma).
}
\label{pion_mu000}
\end{figure}

\begin{figure}
\vspace{-1cm}

\epsfig{figure=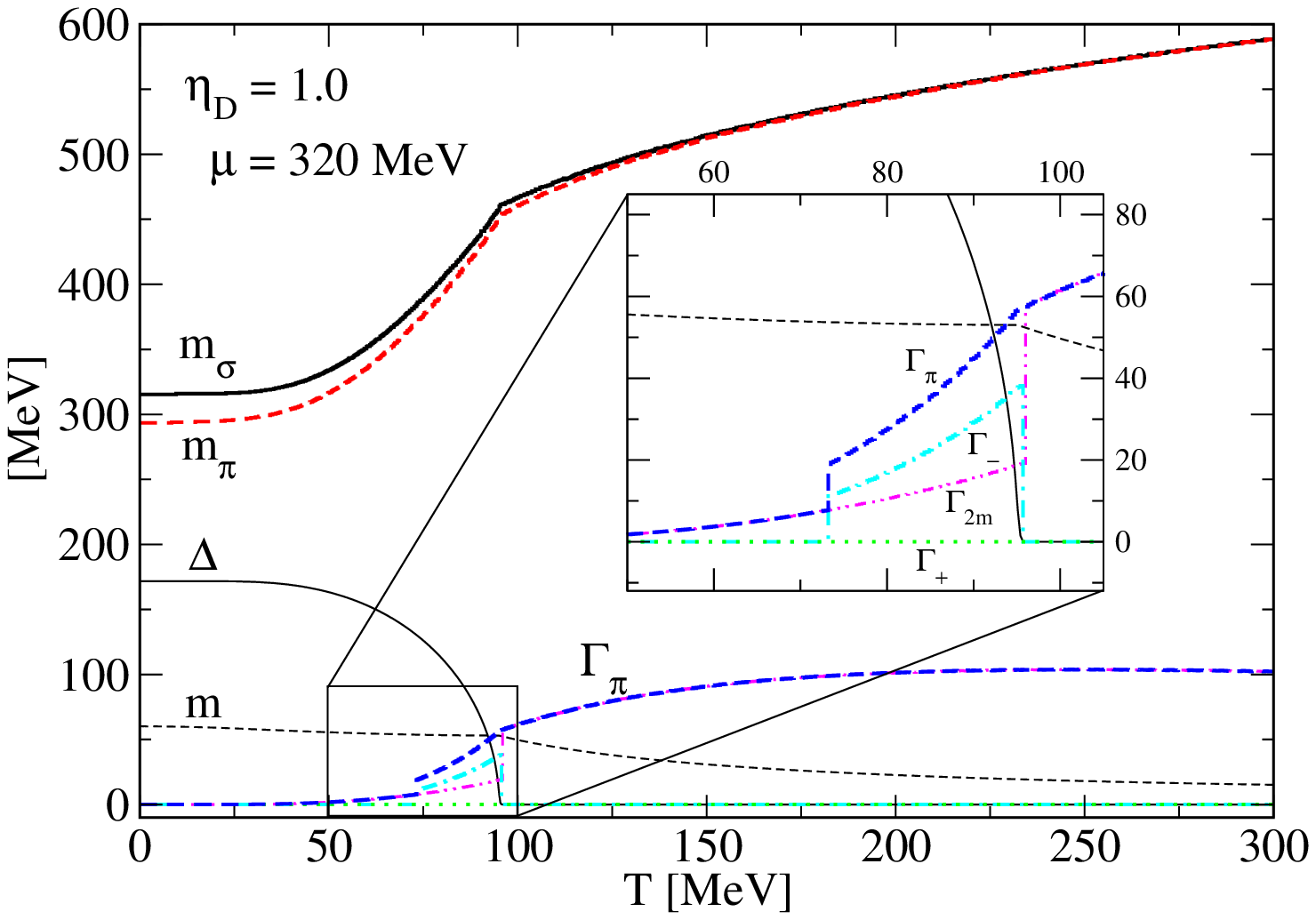,width=0.65\textwidth,angle=-90}
\vspace{-1cm}
\epsfig{figure=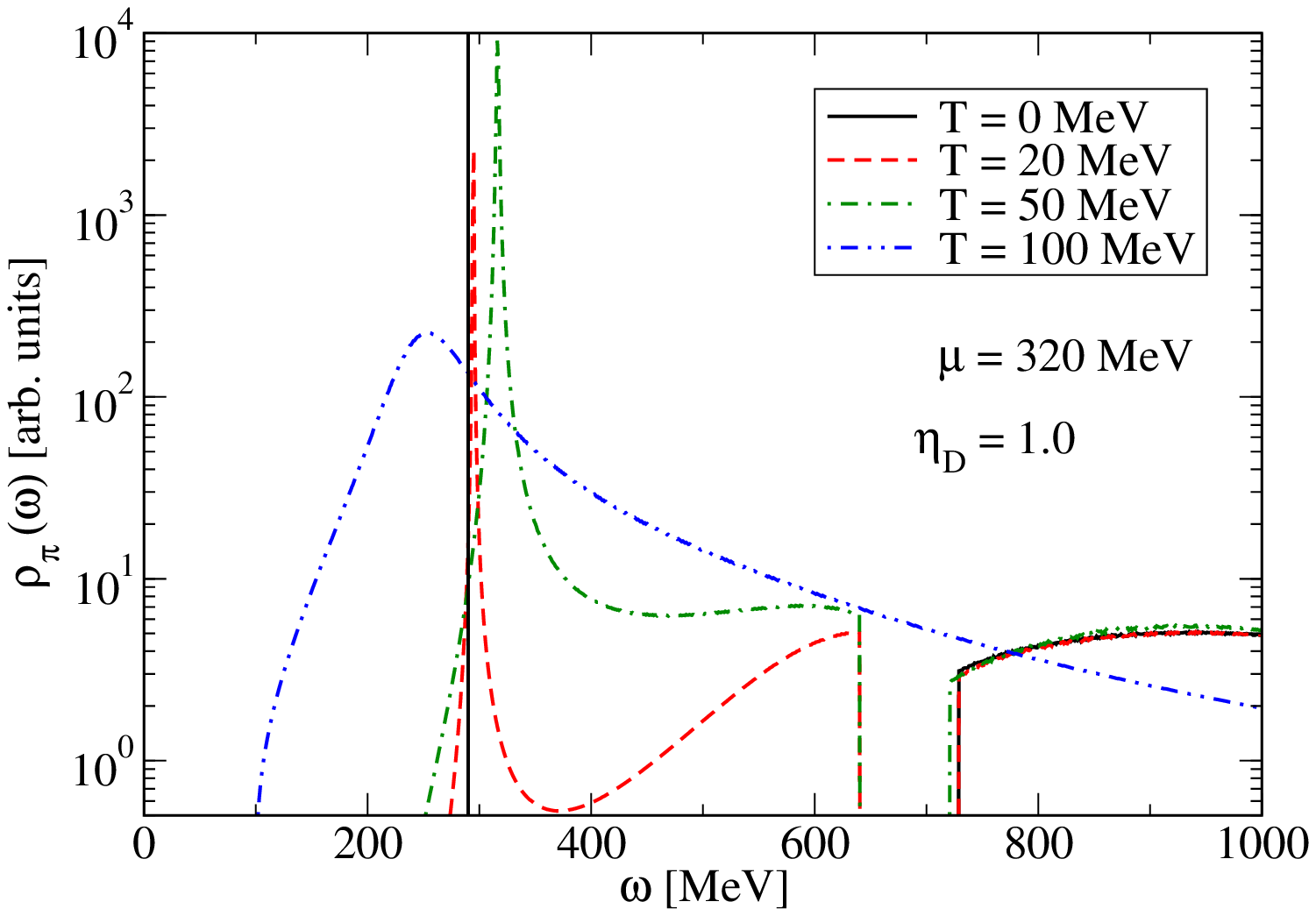,width=0.65\textwidth,angle=-90}
\caption{{\it Upper panel:} 
Mass spectrum of mesons ($\pi$, $\sigma$) as a function 
of the temperature for finite chemical potential $\mu_B=320$ MeV and strong 
diquark coupling $\eta_D=1.0$ in the 2SC phase.
Below the threshold  $E_{\rm th}=2~m_q$ for the onset of the decay width 
$\Gamma_{2m}$ there is another process due to the lower threshold
$E_+ - E_-$ switching on (see inset).
{\it Lower panel:}
Spectral function for pionic correlations for $\mu_B = 320$ in the 2SC phase
at $T = 0$  and at different temperatures around the critical temeprature
for the 2SC phase.
}
\label{pion_mu320}
\end{figure}
Next we want to discuss the pionic excitations in the presence of a 
diquark condensate in the 2SC phase, see Fig.~\ref{pion_mu000}. 
We choose $\mu = 320$ MeV and discuss the effect of melting the 2SC 
diquark condensate by increasing the temperature from $T = 0$ to 
$T > T_c$, where  $T_c = 95$ MeV is the critical temperature for the second 
order transition to the normal quark matter phase. 
We observe the remarkable fact that the 2SC condensate stabilizes the pion 
at $T = 0$ as a true bound state, although the pion mass exceeds by far the 
threshold $2m$. This effect is due to a compensation of gapped and ungapped
quark modes and has been discussed before by Ebert et al.
 \cite{Ebert:2004dr} for $T = 0$ only. Here we extend this study to the 
finite temperature case, where the pion is still a very good resonance, but
obtains a finite width. 
At the critical temperature $T_c$, the normal pion width is 
restored. 
But already before $T = T_c$ is reached, the threshold $X_-$ is reached and 
the corresponding decay process is opened with a considerable width of 
$\mathcal O$(50 MeV).
From the pion spectral function in the lower panel of Fig.~\ref{pion_mu000}
we observe the gap in the excitation spectrum due to the presence of the 
diquark gap. At $T > T_c$, a resonance type spectral function with a threshold
at $\omega = 2m$ and a resonance peak at $\omega \sim 250$ MeV is
obtained. 

The discussion of the mesonic modes in the 2SC phase points to a very rich 
spectrum of excitations which eventually lead to specific new observable
signals of this hypothetical phase. 
The CBM experiment\footnote{contribution by P. Senger} planned at FAIR
Darmstadt  and the NICA project\footnote{contribution by A.S. Sorin et al.} 
at JINR Dubna could be capable of 
creating thermodynamical conditions for the observation of these excitations 
in the experiment. In view of this discovery potential, we want to outline a 
few points for the further development of the theoretical approach.
}
\section{Further Developments}{ 
In this contribution we have described the first steps into the interesting and
very complex physics of the relativistic BEC-BCS crossover theory.
As the next steps following this development, some of the approximations
can be removed.
In particular, one should next
\begin{itemize}
\item evaluate the full spectrum of diquark states,
including their mixing with mesonic channels
\item study the backreaction of the correlations on the meanfield
(selfconsistent meanfield)
\item include higher orders in the one-fermion-loop approximation
(diquark-diquark and diquark-meson interactions)
\item study the effect of the color neutrality condition by adjusting color 
chemical potential(s)
\item study the effect of charge neutrality (gapless superconductivity)
\item evaluate the contribution from diquark-antidiquark annihilation to 
the photon propagator (Maki-Thompson and Aslamasov-Larkin terms).
\end{itemize}
In particular the latter point bears a big potential for applications
to the diagnostics of dense qurk matter formed, e.g., in not too high-
energy nucleus-nucleus collisions.
The onset of color superconductivity not only changes the spectrum of diquark
states (occurrence of the Goldstone bosons) but due to the
nonvanishing diquark gap additional terms for the diquark annihilation
process into the observable dilepton channel arise which stem from the
nonvanishing anomalous propagator contributions.
Since the critical temperature for the color superconductivity transition 
might be as high as 100 MeV there is a fair chance to observe traces of
this transition with the future CBM experiment at FAIR Darmstadt.
}
\section{Conclusions}{
We have derived the gap equations and Bethe-Salpeter equations for
the simultaneous treatment of quantized density fluctuations
(particle-antiparticle modes = mesons) in RPA approximation (one-fermion-loop
polarization function) and quantized pairing fluctuations (particle-particle
modes = diquarks) in ladder approximation
(no crossed-ladder or vertex corrections) within the path integral
formalism, i.e. a fully relativistic field theoretical treatment. 
For the a priori
unknown interaction, a local current-current coupling (NJL model) has been used
which in the nonrelativistic limit becomes equivalent to the BCS model of
superconductivity. %
After fixing the parameters of the model to the light meson spectrum
in the vacuum, the diquark coupling remains as a free parameter which
has been used to extend the model beyond the traditional range of applications
into the region of BEC-BCS crossover, where both diquark
bound states and scattering states occur simultaneously and determine
the physical properties of the system. %
Recently, the tuning of the coupling in (nonrelativistic) low-temperature 
systems of fermionic atoms with Feshbach resonances
in an external magnetic field could be used to investigate the BEC-BCS 
crossover in the laboratory.
The nonrelativistic limit of the present approach can be used to interprete 
such results within a local coupling model. The fully relativistic form can 
be applied to model strong coupling QCD and
to investigate the effects of finite temperature and density on
the phase structure. %
We have presented in this contribution the phase diagram of quark matter at
strong and very strong coupling, delineating a possible region of BEC-BCS
crossover.
The origin of the  BEC-BCS crossover in superconducting quark matter is the 
Mott transition for diquark bound states. 
We explain the physics of the Mott transition on the example of mesonic 
correlations.
We investigate the spectral function for pionic correlations
(bound and scattering states) outside ($T > T_c$) and for the first time also
inside  ($T < T_c$) the color superconductivity region.
We find the thresholds for the dissociation of pionic bound states
into unbound, but resonant scattering states in the quark-\-antiquark 
continuum.
For the lifetime of pionic resonances in the quark matter phase diagram see
\cite{Zablocki:2008sj}. %
Summarizing, we have provided the framework for a study of hot and dense
fermion matter beyond the mean field within a fully relativistic approach, with
yet local interaction and in Gaussian approximation.
The applications to QCD matter within a NJL model provide us with a phase 
diagram for quark matter which is now augmented with information about
the presence of strong mesonic and diquark correlations with possible 
consequences in the phenomenology of relativistic heavy-ion collisions and the 
interiors of compact stars.
}

\end{document}